\newcommand{\be}{\begin{equation}}  
\newcommand{\ee}{\end{equation}}  
\newcommand{\beq}{\begin{eqnarray}}  
\newcommand{\eeq}{\end{eqnarray}}
\documentclass{PoS}

\usepackage{amsmath}
\usepackage{graphicx}

\title{Investigation of the tetraquark candidate $a_0(980)$: \\ technical aspects and preliminary results}

\ShortTitle{Investigation of the tetraquark candidate $a_0(980)$}

\author{\speaker{Abdou Abdel-Rehim}${}^{1}$, Constantia Alexandrou${}^{1,2}$, \speaker{Joshua Berlin}${}^{3}$, Mattia Dalla Brida${}^{4}$, Mario Gravina${}^{5}$, Marc Wagner${}^{3}$\\

\email{a.abdel-Rehim@cyi.ac.cy},
\email{alexand@ucy.ac.cy},
\email{berlin@th.physik.uni-frankfurt.de},
\email{mattia.dallabrida@gmail.com},
\email{mario.gravina@fis.unical.it},
\email{mwagner@th.physik.uni-frankfurt.de}\\
        
        ${}^{1}$Computation-based Science and Technology Research Center, The Cyprus Institute, \\ $\quad$ 20 Kavafi Street, 2121 Nicosia, Cyprus\\
        ${}^{2}$Department of Physics, University of Cyprus, P.O. Box 20537, 1678 Nicosia, Cyprus\\
        ${}^{3}$Goethe-Universit\"at Frankfurt am Main, Institut f\"ur theoretische Physik, \\ $\quad$ Max-von-Laue-Stra{\ss}e 1, D-60438 Frankfurt am Main, Germany\\
        ${}^{4}$School of Mathematics, Trinity College Dublin, Dublin 2, Ireland,\\
		$\quad$ \& NIC, DESY, Platanenallee 6, 15738 Zeuthen, Germany\\
        ${}^{5}$Universit\`a della Calabria, Via Pietro Bucci, 87036 Arcavacata di Rende Cosenza, Italy \\
        }

\abstract{
We discuss technical aspects and first results of a lattice QCD study of the  $a_0(980)$ state. We employ various interpolating operators of quark-antiquark, mesonic molecule, diquark-antidiquark and two-meson type. Both connected and disconnected contributions including diagrams with closed fermion loops are computed. To keep statistical errors small, it is essential to optimize the computation of these diagrams by choosing that combination of techniques most appropriate for each type of diagram from the correlation matrix of interpolating operators. We illustrate, how this can be done, by discussing certain diagrams in detail. We also present preliminary results corresponding to a $4\times 4$ submatrix computed with 2+1 flavors of clover fermions.
}

\FullConference{The 32nd International Symposium on Lattice Field Theory\\
		 23-28 June, 2014\\
		 Columbia University New York, NY}

\begin{document}


\section{Introduction}

\begin{figure}[htb]
\begin{center}
\includegraphics[width=4.5cm]{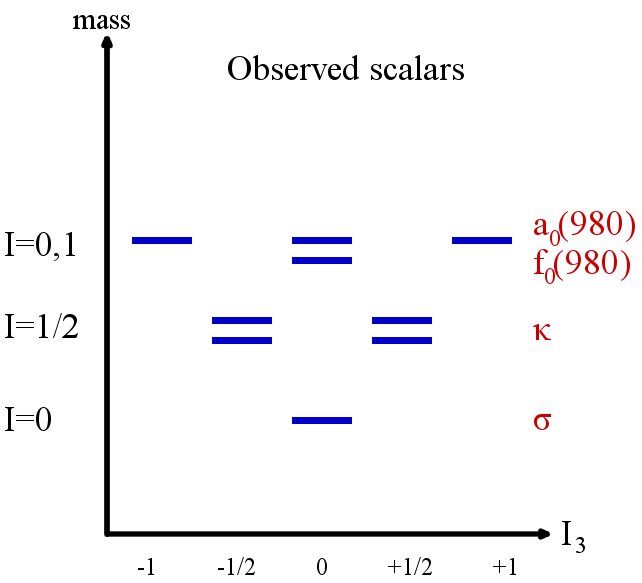}\hspace{.2cm}
\includegraphics[width=4.5cm]{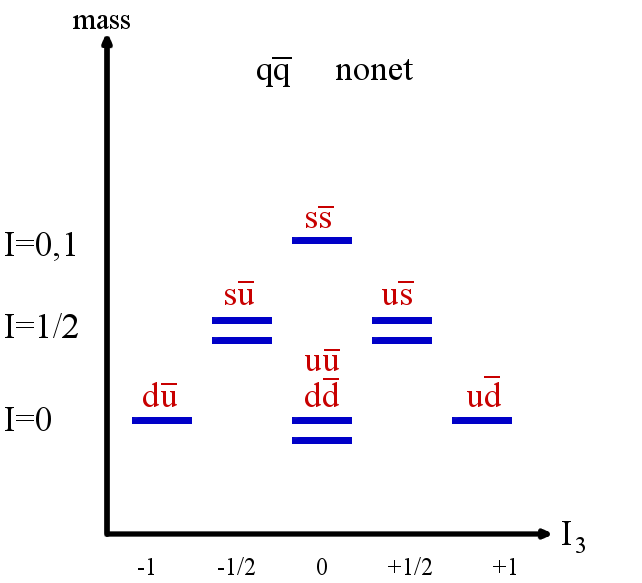}\hspace{.2cm}
\includegraphics[width=4.5cm]{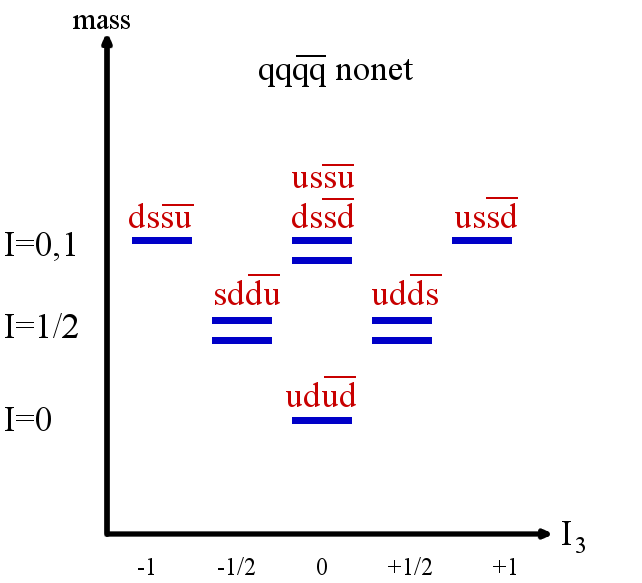} 
\end{center}
\caption{\label{fig:masspattern}The spectrum of light scalar mesons ($J^P=0^+$).
From left to right:
\textbf{(a)}~Experimental results.
\textbf{(b)}~Conventional quark model (quark-antiquark structure).
\textbf{(c)}~Assuming a four-quark structure.
}
\end{figure}

Our understanding  of the light scalar meson sector ($J^P = 0^+$) is incomplete \cite{Pelaez:2014rla,Pelaez:2013jp,pdg:sep2014}. The observed mass ordering of the states $\sigma$, $f_0(980)$, $\kappa$ and $a_0(980)$ appears inverted from what would be naively expected from conventional quark models (cf.\ Fig.~\ref{fig:masspattern}a and Fig.~\ref{fig:masspattern}b). Using a single quark and a single antiquark isospin $I = 1$ can only be realized with two light quarks, whereas for $I = 0$ either two light quarks or two strange quarks are possible. Thus in the conventional quark model the flavor structure of these scalar mesons would be the following SU(3) flavor nonet \cite{Jaffe:2004ph}:
\begin{eqnarray}
\nonumber & & \hspace{-0.7cm} I = 0\phantom{/2} \quad \rightarrow \quad \sigma = \frac{1}{\sqrt{2}} \Big(u \bar{u} + d \bar{d}\Big) \ \ , \ \ f_0 = s \bar{s} \\
\nonumber & & \hspace{-0.7cm} I = 1/2 \quad \rightarrow \quad \kappa^+ = u \bar{s} \ \ , \ \ \kappa^0 = d \bar{s} \ \ , \ \ \bar{\kappa}^0 = s \bar{d} \ \ , \ \ \kappa^- = s \bar{u} \\
\label{qq} & & \hspace{-0.7cm} I = 1\phantom{/2} \quad \rightarrow \quad a_0^+ = u \bar{d} \ \ , \ \ a_0^0 = \frac{1}{\sqrt{2}} \Big(u \bar{u} - d \bar{d}\Big) \ \ , \ \ a_0^- = d \bar{u} .
\end{eqnarray}
Moreover, such an assignment does not explain the mass degeneracy of $f_0(980)$ and $a_0(980)$ and it is hard to understand, why $\sigma$ and $\kappa$ are broader than $f_0(980)$ and $a_0(980)$.

Alternatively one can assume a four-quark structure with quark content
\begin{eqnarray}
\nonumber & & \hspace{-0.7cm} I = 0\phantom{/2} \quad \rightarrow \quad \sigma = u d \bar{u} \bar{d} \ \ , \ \ f_0 = \frac{1}{\sqrt{2}} \Big(u s \bar{u} \bar{s} + d s \bar{d} \bar{s}\Big) \\
\nonumber & & \hspace{-0.7cm} I = 1/2 \quad \rightarrow \quad \kappa^+ = u d \bar{d} \bar{s} \ \ , \ \ \kappa^0 = u d \bar{u} \bar{s} \ \ , \ \ \bar{\kappa}^0 = u s \bar{u} \bar{d} \ \ , \ \ \kappa^- = d s \bar{u} \bar{d} \\
\label{4q} & & \hspace{-0.7cm} I = 1\phantom{/2} \quad \rightarrow \quad a_0^+ = u s \bar{d} \bar{s} \ \ , \ \ a_0^0 = \frac{1}{\sqrt{2}} \Big(u s \bar{u} \bar{s} - d s \bar{d} \bar{s}\Big) \ \ , \ \ a_0^- = d s \bar{u} \bar{s} .
\end{eqnarray}
Within this interpretation both the mass degeneracy of $f_0(980)$ and $a_0(980)$ and the mass ordering of the whole nonet is simple to understand (cf.\ Fig.~\ref{fig:masspattern}c). The rather large width of $\sigma$ and $\kappa$ is also easier to explain, since the decay channels to $\pi + \pi$ and $K + \pi$, respectively, are OZI allowed. 

A number of lattice QCD studies of light scalar mesons have been published in the last couple of years \cite{Bernard:2007qf,Gattringer:2008be,Prelovsek:2008qu,Liu:2008ee,Wakayama:2012ne,Prelovsek:2013ela}. In this work we continue our investigation of the light scalar nonet \cite{Daldrop:2012sr,Alexandrou:2012rm,Wagner:2012ay,Wagner:2013nta,Wagner:2013jda,Wagner:2013vaa} focusing on the study of the $a_0(980)$ state. We use a variety of interpolating operators with the aim to shed some light on the structure of the $a_0(980)$. The interpolators include a conventional quark-antiquark operator as well as different types of four quark operators with mesonic molecule, diquark-antidiquark and two-meson structure. In the corresponding correlation matrix several diagrams with disconnected pieces or with closed fermion loops are present, which are particularly difficult to compute. To obtain an acceptable signal-to-noise ratio, it is imperative to identify the most efficient strategy of computation for each diagram. In section~\ref{SEC012} and section~\ref{sec:tfteotcme} we will describe these technical aspects in detail. In section~\ref{SEC002} the lattice setup is discussed and first numerical results are presented.


\section{\label{SEC012}Interpolating operators and correlation matrix}

Our variational basis of interpolating operators $\mathcal{O}^j$ entering the correlation matrix
\begin{eqnarray}
C_{j k}(t) = \Big\langle \mathcal{O}^j(t) \mathcal{O}^{k \dag}(0) \Big\rangle
\end{eqnarray}
is the following:
\begin{eqnarray}
\mathcal{O}^1 = \mathcal{O}^{q\bar q} &=& \sum_{\bf{x}} \Big( {\bar d}_{\bf x} {u}_{\bf x} \Big) \label{eq:operatorone}
 \\
\mathcal{O}^2 = \mathcal{O}^{K \bar{K} \text{, point}} &=& \sum_{\bf{x}} \Big( {\bar s}_{\bf x} \gamma_5 {u}_{\bf x} \Big) \Big( {\bar d}_{\bf x} \gamma_5 {s}_{\bf x} \Big) \label{eq:operatortwo}
\\
\mathcal{O}^3 = \mathcal{O}^{\eta_s \pi \text{, point}} &=& \sum_{\bf{x}} \Big( {\bar s}_{{\bf x}} \gamma_5 {s}_{{\bf x}} \Big) \Big( {\bar d}_{{\bf x}} \gamma_5 {u}_{{\bf x}} \Big)
 \\
\mathcal{O}^4 = \mathcal{O}^{Q \bar Q} &=& \sum_{\bf{x}} \epsilon_{abc} \Big( {\bar s}_{{\bf x}, b} {(C \gamma_5)} {\bar d}^T_{{\bf x}, c} \Big) \epsilon_{ade} \Big( {u}^T_{{\bf x}, d} {(C \gamma_5)} s_{{\bf x}, e} \Big) \label{eq:operatorfour}
\\
\mathcal{O}^5 = \mathcal{O}^{K\bar{K} \text{, 2-part}} &=& \sum_{{\bf x,y}} \Big( {\bar s}_{{\bf x}} \gamma_5 {u}_{{\bf x}} \Big) \Big( {\bar d}_{{\bf y}} \gamma_5 {s}_{{\bf y}} \Big) \label{eq:operatorfive}
\\
\mathcal{O}^6 = \mathcal{O}^{\eta_s \pi \text{, 2-part}} &=& \sum_{\bf{x,y}} \Big( {\bar s}_{{\bf x}} \gamma_5 {s}_{{\bf x}} \Big) \Big( {\bar d}_{{\bf y}} \gamma_5 {u}_{{\bf y}} \Big), \label{eq:operatorsix}
\end{eqnarray}
where $C$ is the charge conjugation matrix. The first interpolating operator $\mathcal{O}^{q\bar q}$ is the conventional quark-antiquark ``quark model interpolator''. Since the rest of the interpolating operators have four-quarks, the off-diagonal elements $C_{1,j}$ and $C_{j,1}$, $j=2,\cdots,6$ of the correlation matrix will have closed fermion loops or propagators that start and end on the same timeslice. The interpolating operators $\mathcal{O}^{K \bar{K} \text{, point}}$, $\mathcal{O}^{\eta_s \pi \text{, point}}$ and $\mathcal{O}^{Q \bar Q}$ are  four-quark operators with all quark fields located on the same point in space. The first two have a mesonic molecule structure ($K \bar{K}$ and $\eta_s \pi$), whereas the third has diquark-antidiquark structure (here we use the lightest diquark and antidiquark corresponding to spin structure $C \gamma_5$ \cite{Jaffe:2004ph,Alexandrou:2006cq,Wagner:2011fs}) and is expected to have a large overlap with a possibly existing tetraquark state. The last two interpolating operators $\mathcal{O}^{K\bar{K} \text{, 2-part}}$ and $\mathcal{O}^{\eta_s \pi \text{, 2-part}}$ are also made of two mesons, but each meson has been projected to zero momentum, i.e.\ their positions are independent from each other. 

\begin{figure}[htb]
\begin{center}
\vspace{-1cm}
\hspace{-5cm}
\begin{minipage}{0.5\textwidth}
\includegraphics[width=11.cm, page=1]{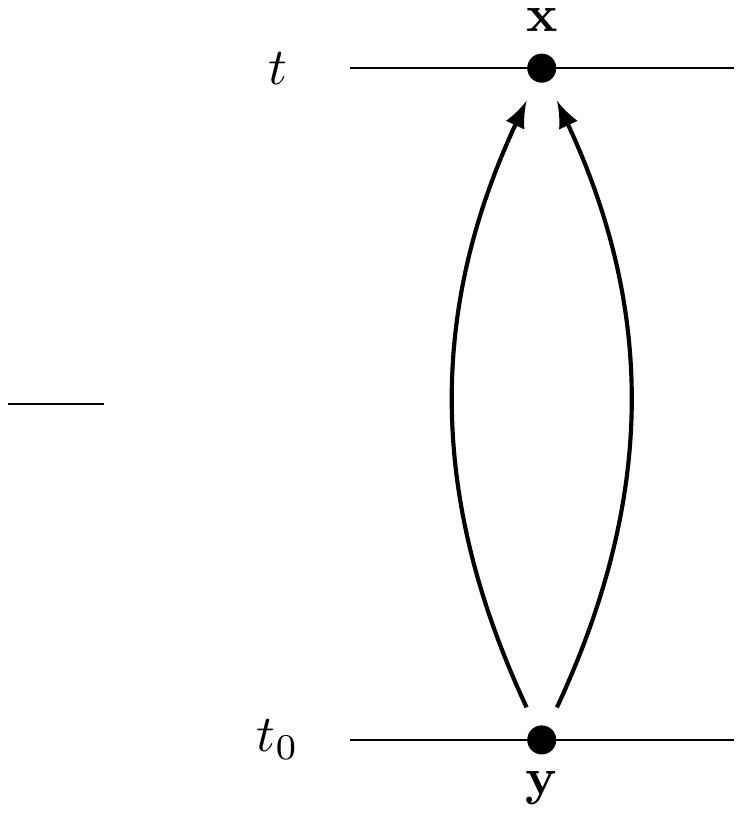}
\end{minipage}
\begin{minipage}{0.5\textwidth}
\includegraphics[width=11.cm, page=8]{diagrams.pdf} 
\end{minipage}
\end{center}
\vspace{-2.5cm}
\caption{\label{fig:diagramintro}Diagrammatic representation of correlation matrix elements.
From left to right:
\textbf{(a)}~$C_{1 1}$, eq.\ \protect\eqref{eq:A}.
\textbf{(b)}~$C_{2 2}$ and $C_{3 3}$, eq.\ \protect\eqref{eq:B} and eq.\ \protect\eqref{eq:C}.
}
\end{figure}

The matrix elements $C_{j k}$ can be expressed in terms of quark propagators and represented diagrammatically. For example the $C_{1 1}$ matrix element is given by
\begin{eqnarray}
C_{1 1}(t) = 
\Big\langle \mathcal{O}^{q \bar q}(t) \mathcal{O}^{q \bar q ^\dag}(0) \Big\rangle =  
- \sum_{\mathbf{x},\mathbf{y}} \Big\langle {\rm Tr}\Big( \gamma_5 G^d({\bf x},t;{\bf y},0)^\dag \gamma_5 G^u({\bf x},t;{\bf y},0) \Big) \Big\rangle , \label{eq:A}
\end{eqnarray}
where $G^{u/d}$ denotes the $u/d$ propagator and the trace is over the spin and color components. This expression corresponds to the diagram shown in Fig.~\ref{fig:diagramintro}a. Similarly one can write the matrix elements $C_{2 2}$ and $C_{3 3}$ as
\begin{eqnarray}
\nonumber & & \hspace{-0.7cm} C_{2 2}(t) = \Big\langle \mathcal{O}^{K \bar K \text{, point}}(t) {\mathcal{O}^{K \bar K \text{, point}}}(0)^\dag \Big\rangle = \\
\nonumber & & = \sum_{\bf x,y}
\Big\langle 
{\rm Tr}\Big({G^s}({\bf x}, t; {\bf y}, 0)^\dag G^u({\bf x}, t; {\bf y}, 0)\Big)
{\rm Tr}\Big({G^d}({\bf x}, t; {\bf y}, 0)^\dag G^s({\bf x}, t; {\bf y}, 0)\Big) 
\Big\rangle \\
\label{eq:B} & & \hspace{0.675cm} -\sum_{\bf x,y}
\Big\langle 
{\rm Tr}\Big(\gamma_5 G^u({\bf x}, t; {\bf y}, 0) \gamma_5 G^s({\bf y}, 0; {\bf y}, 0) {G^d}({\bf x}, t; {\bf y}, 0)^\dag G^s({\bf x}, t; {\bf x}, t)\Big)
\Big\rangle \\
\nonumber & & \hspace{-0.7cm} C_{33}(t) = \Big\langle \mathcal{O}^{\eta_s \pi \text{, point}}(t) {\mathcal{O}^{\eta_s  \pi \text{, point}}}(0)^\dag \Big\rangle = \\
\nonumber & & = \sum_{\bf x,y}
\Big\langle
{\rm Tr}\Big({G^s}({\bf x}, t; {\bf y}, 0)^\dag G^s({\bf x}, t; {\bf y}, 0)\Big) 
{\rm Tr}\Big({G^d}({\bf x}, t; {\bf y}, 0)^\dag G^u({\bf x}, t; {\bf y}, 0)\Big) 
\Big\rangle \\
\label{eq:C} & & \hspace{0.675cm} -\sum_{\bf x,y}
\Big\langle
{\rm Tr}\Big(\gamma_5 G^s({\bf y}, 0; {\bf y}, 0)\Big) {\rm Tr}\Big({G^d}({\bf x}, t; {\bf y}, 0)^\dag G^u({\bf x}, t; {\bf y}, 0)\Big) {\rm Tr}\Big(\gamma_5 G^s({\bf x}, t; {\bf x}, t)\Big)
\Big\rangle ,
\end{eqnarray}
where the second terms in \eqref{eq:B} and \eqref{eq:C} contain closed fermion loops. In our diagrammatic language, which displays the spacetime structure, but does not take into account spin and color indices, $C_{2 2}$ and $C_{3 3}$ are represented by the same diagram (cf.\ Fig.~\ref{fig:diagramintro}b).

\begin{figure}[htb]
\begin{center}
\includegraphics[width=15.cm,page=2]{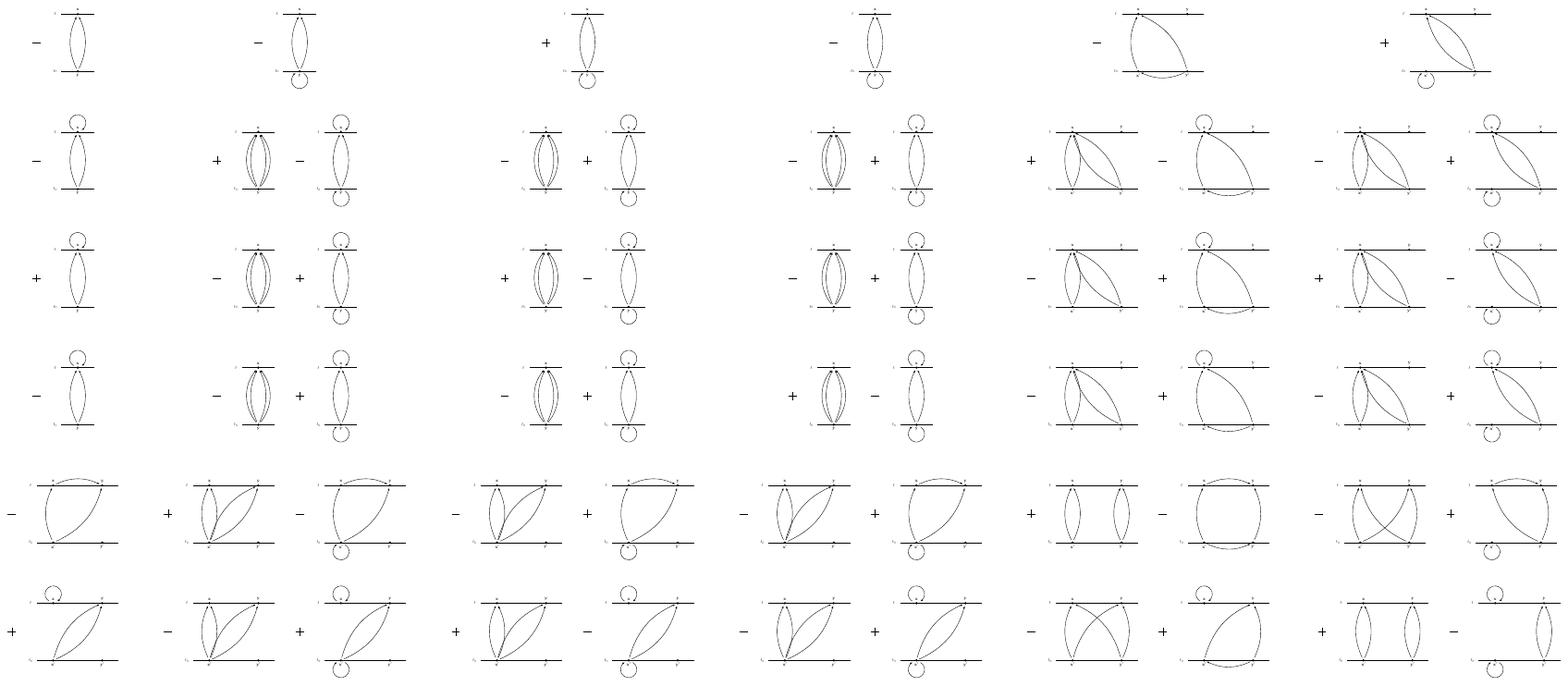}
\end{center}
\vspace{-3cm}
\caption{\label{fig:diagrammatrix}Diagrammatic representation of the $6 \times 6$ correlation matrix $C_{j k}$ corresponding to the interpolating operators \protect\eqref{eq:operatorone} to \protect\eqref{eq:operatorsix}.}
\end{figure}

After applying this procedure to all matrix elements of the $6 \times 6$ correlation matrix $C_{j k}$ one arrives at the diagrammatic matrix representation shown in Fig.~\ref{fig:diagrammatrix}. One can easily see the necessity of computing closed fermion loops and timeslice propagators, e.g.\ for the correlations of $\mathcal{O}^{q\bar q}$ and four-quark interpolating operators (first row and first column).

\section{\label{sec:tfteotcme}Techniques to compute the correlation matrix elements}

In this section we discuss and compare different methods to compute propagators and the correlation matrix elements $C_{j k}$. For each diagram one should choose the optimal combination of techniques in a sense that the required CPU time is minimized and/or the signal-to-noise ratio is maximized.

There are four ways to compute quark propagators and correlators, which we try to combine most efficiently for each diagram:
\begin{enumerate}
\item[a)] \textbf{Fixed-source propagators:} (for details cf.\ e.g.\ \cite{DeGrand:2006zz,Gattringer:2010zz}) \\ A straightforward method to compute propagators $G(x;x_0)$ from a single point in spacetime $x_0$ to any other point in spacetime $x$ (therefore, also called point-to-all propagators). Fixed-source propagators are prohibitively expensive, when a diagram involves sums over both ends of the propagator, i.e.\ $\sum_{\mathbf{x},\mathbf{x}_0}$.

\item[b)] \textbf{Stochastic timeslice-to-all propagators:} (for details cf.\ e.g.\ \cite{Bernardson:1993he,Dong:1993pk}) \\ Using stochastic $Z(2)\times Z(2)$ noise on a single timeslice $t_0$ one can stochastically estimate propagators $G(x;\mathbf{x}_0,t_0)$ from any point in space $\mathbf{x}_0$ at time $t_0$ to any other point in spacetime $x$. This technique is particularly useful, when computing closed fermion loops, i.e.\ diagrams involving $\sum_\mathbf{x} G(\mathbf{x},t;\mathbf{x},t)$.

\item[c)] \textbf{One-end trick:} (for details cf.\ e.g.\ \cite{Foster:1998vw,McNeile:2006bz}) \\ An efficient method to stochastically estimate a pair of propagators combined by a spatial sum $\sum_{\mathbf{x}_0}$ at time $t_0$, i.e.\ expressions containing $\sum_{\mathbf{x}_0} G(x;\mathbf{x}_0,t_0) G^\dagger(\mathbf{x}_0,t_0;y)$. If more than two propagators are involved in the sum, the one-end trick is not applicable.

\item[d)] \textbf{Sequential propagators:} (cf.\ e.g.\ \cite{Martinelli:1988rr}) \\ Another possibility to compute a pair of propagators combined by a spatial sum $\sum_{\mathbf{x}_0}$ at time $t_0$, i.e.\ expressions containing $\sum_{\mathbf{x}_0} G(x;\mathbf{x}_0,t_0) G^\dagger(\mathbf{x}_0,t_0;y)$, which does not necessarily involve stochastic sources. Again, if more than two propagators are involved in the sum, it is not possible to compose two of them into a sequential propagator.
\end{enumerate}
These techniques can be combined in many different ways. An example is the combination of a sequential propagator d) and the one-end trick c) to compute the ``triangular diagram'' $C_{1 5}$ as sketched in the following:
\begin{eqnarray}
C_{1 5} = -\Big\langle \sum_{{\bf x}} \underbrace{\sum_{{\bf x'}} {\gamma_5}_{\alpha \beta} G^u_{\beta \gamma}({\bf x}, t;{\bf x'}, t') {\gamma_5}_{\gamma \delta} \underbrace{\sum_{{\bf y'}} G^s_{\delta \epsilon}({\bf x'},t';{\bf y'}, t') G^d_{\epsilon \alpha}({\bf y'}, t';{\bf x}, t)}_{\textrm{sequential propagator } G_{\delta \alpha}^\textrm{sequential}({\bf x'}, t';{\bf x}, t)}}_{\textrm{one-end trick}} \Big\rangle .
\end{eqnarray}
The sequential propagator replaces two of the quark propagators $G^s_{\delta \epsilon}({\bf x'},t';{\bf y'}, t')$ and $G^d_{\epsilon \alpha}({\bf y'}, t';{\bf x}, t)$. Then the one-end trick is used to contract it properly with the third quark propagator $G^u_{\beta \gamma}({\bf x}, t;{\bf x'}, t')$. This approach yields a much better signal-to-noise ratio than e.g.\ combining three stochastic time\-slice-to-all propagators. Another diagram, for which this strategy is very efficient, is the ``rectangular diagram'' in $C_{5 5}$.


\subsection{Selecting the optimal method}

\begin{figure}[htb]
\begin{center}
\includegraphics[width=11.cm]{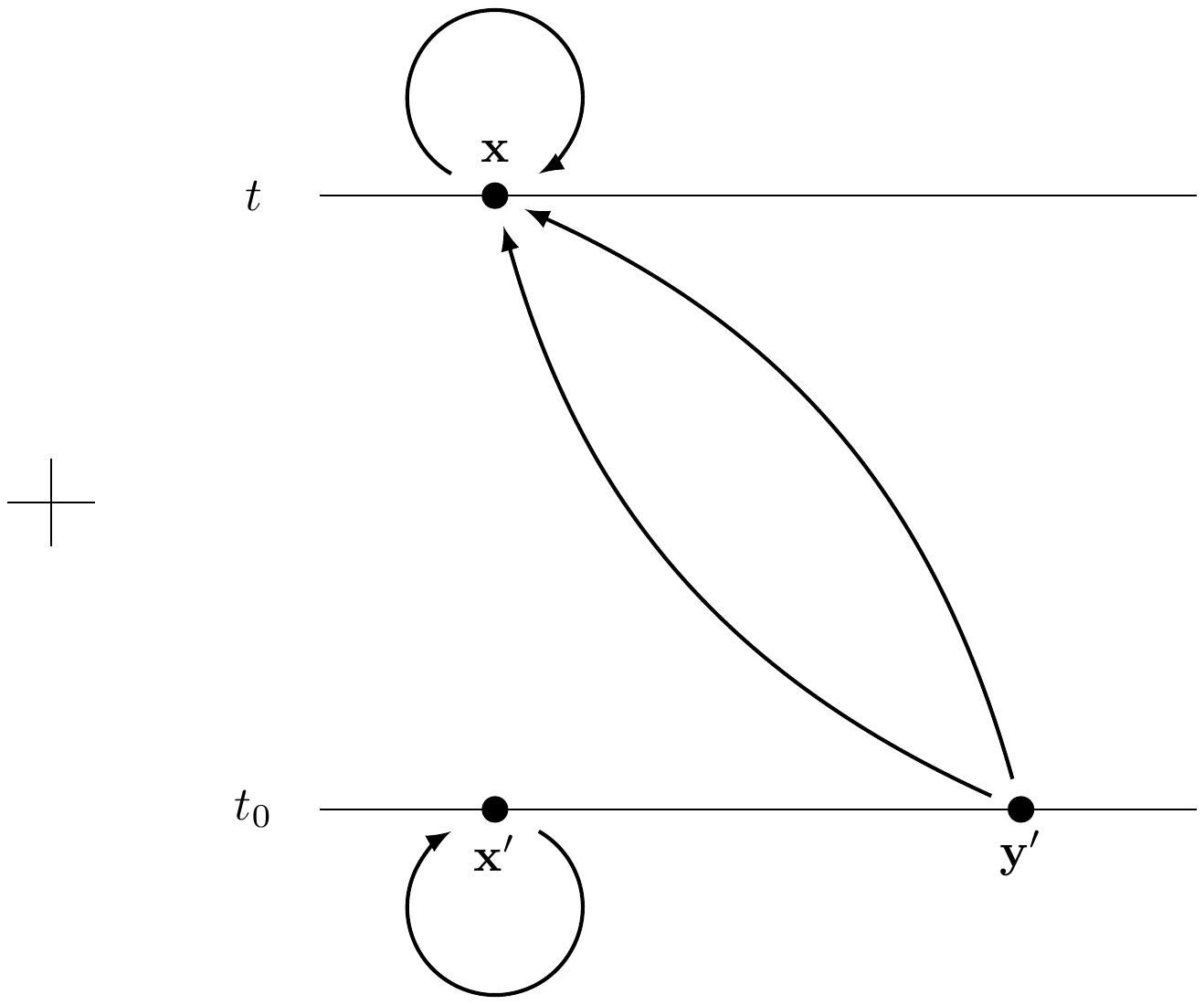}
\end{center}
\vspace{-2.5cm}
\caption{\label{fig:multiplemethods}Diagram contributing to $C_{4 6}$.}
\end{figure}

Each diagram in Fig.~\ref{fig:diagrammatrix} can be computed in a variety of ways. A priori it is usually not clear, which combination of techniques a), b), c) and d) is most efficient. Consider for example the diagram with the two closed fermion loops contributing to $C_{4 6}$ (cf.\ Fig.~\ref{fig:multiplemethods}). Options to evaluate this diagram numerically include the following:
\begin{itemize}
\item[(1)] Compute three fixed-source propagators and a stochastic propagator (fixed-source at $\mathbf{x}$, stochastic propagator for the disconnected loop at $\mathbf{x}'$).

\item[(2)] Use the one-end trick at $\mathbf{y}'$ and a fixed-source propagator for the loop at $\mathbf{x}$ to compute the big connected piece, use a stochastic propagator for the disconnected loop at $\mathbf{x}'$).

\item[(3)] Use the one-end trick at $\mathbf{y}'$ and a stochastic propagator for the loop at $\mathbf{x}$ to compute the big connected piece, use a fixed-source propagator for the disconnected loop at $\mathbf{x}'$).

\item[(4)] Use the one-end trick at $\mathbf{y}'$ and a stochastic propagator for the loop at $\mathbf{x}$ to compute the big connected piece, use another stochastic propagator for the disconnected loop at $\mathbf{x}'$).
\end{itemize}
The last option (4) involves two stochastic propagators. In general increasing the number of stochastic propagators quickly leads to a poor signal-to-noise ratio. Therefore (4) is not expected to be an efficient method. Using a single stochastic propagator and the one-end trick (options (2) and (3)) might still be comparable to option (1), where also a stochastic propagator is involved. It probably needs an exploratory numerical study before (2) and (3) should be discarded. Whether (2) and (3) perform on a similar level or one is superior to the other, is also not obvious. Probably option (1) is the most efficient choice, since only one of the four propagators is treated by stochastic methods.


\subsection{\label{SEC010}An example of a numerical comparison of methods}

\begin{figure}[htb]
\begin{center}
\includegraphics[width=7.5cm, page=2]{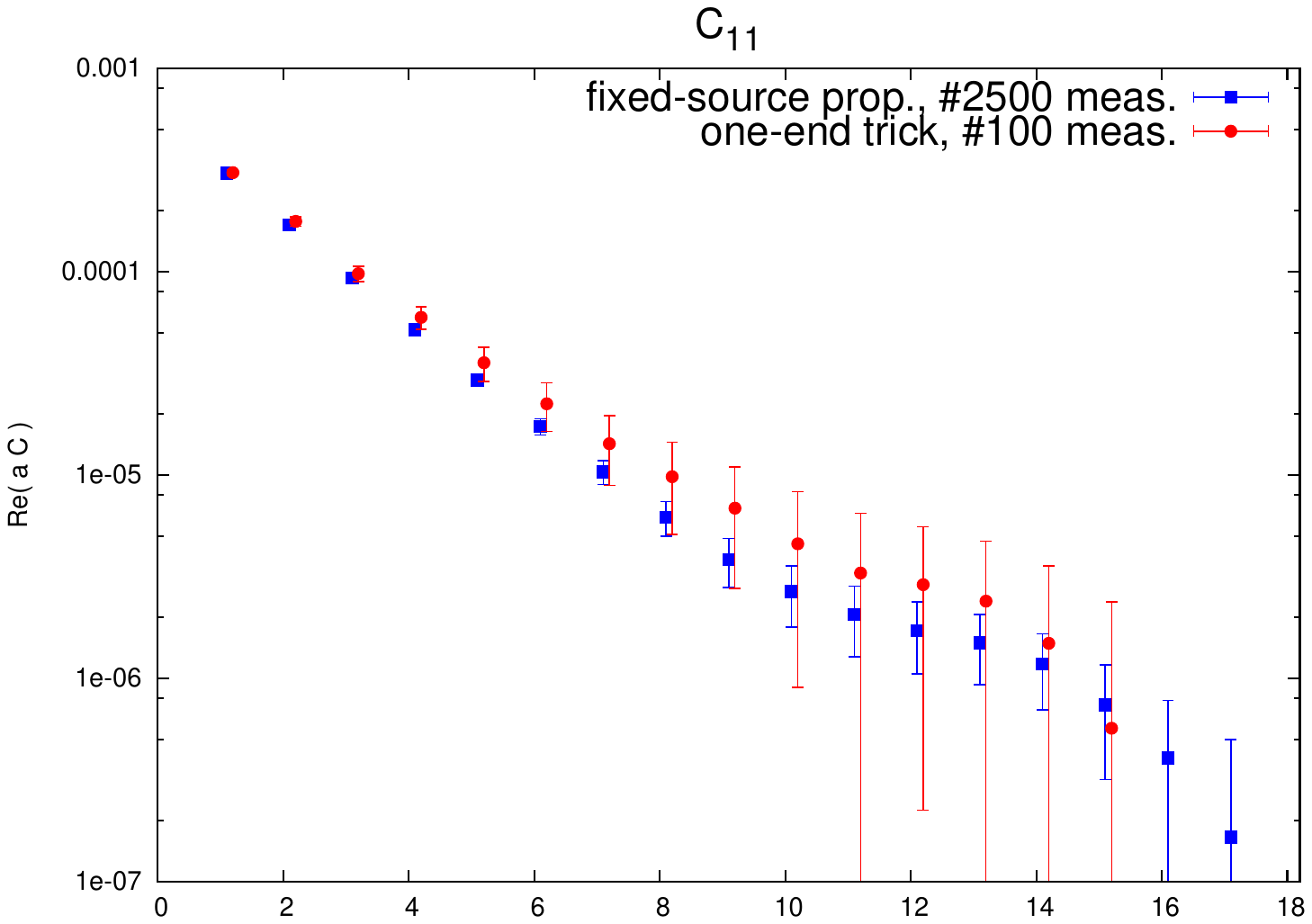}
\includegraphics[width=7.5cm, page=3]{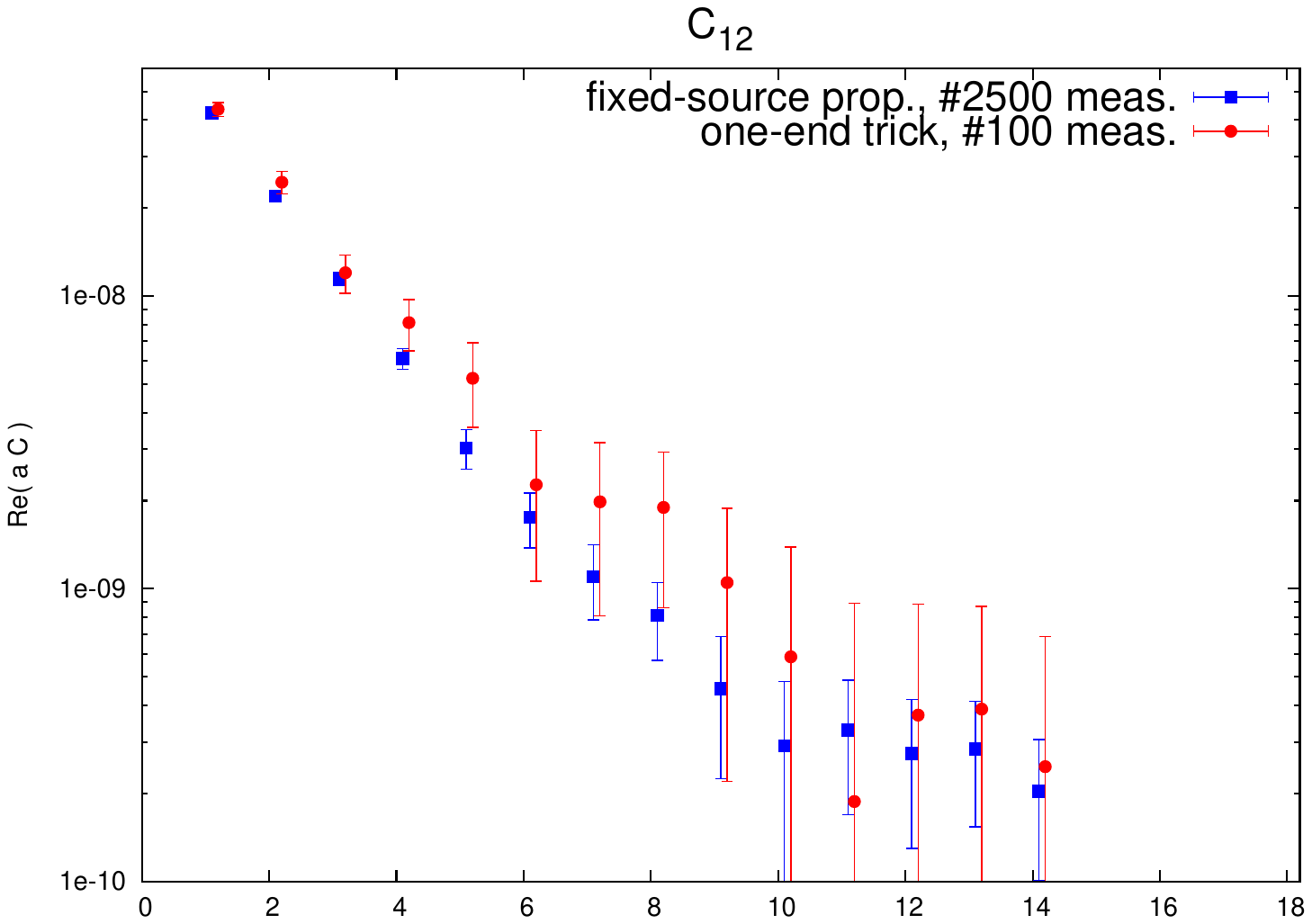}
\includegraphics[width=7.5cm, page=5]{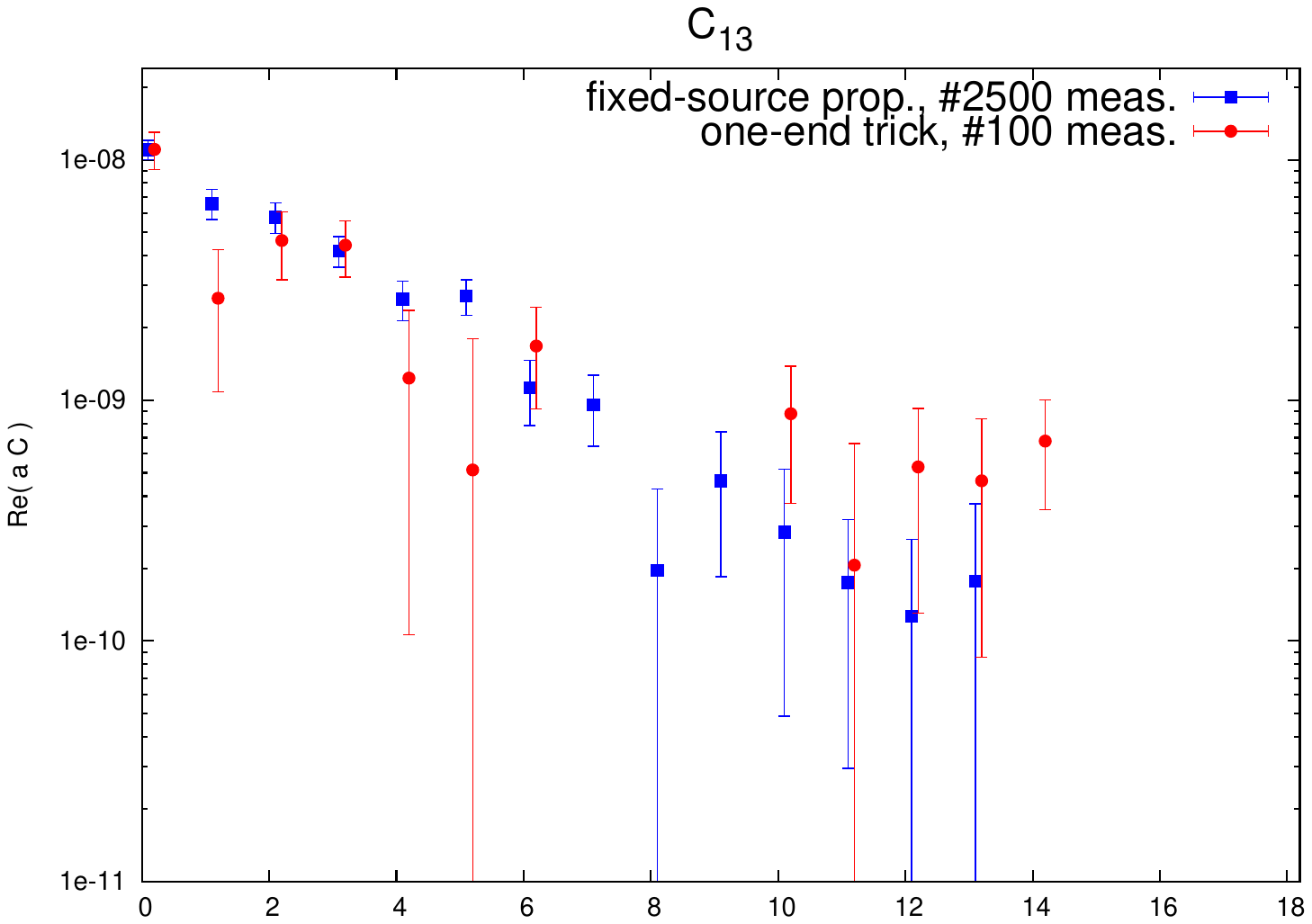}
\includegraphics[width=7.5cm, page=7]{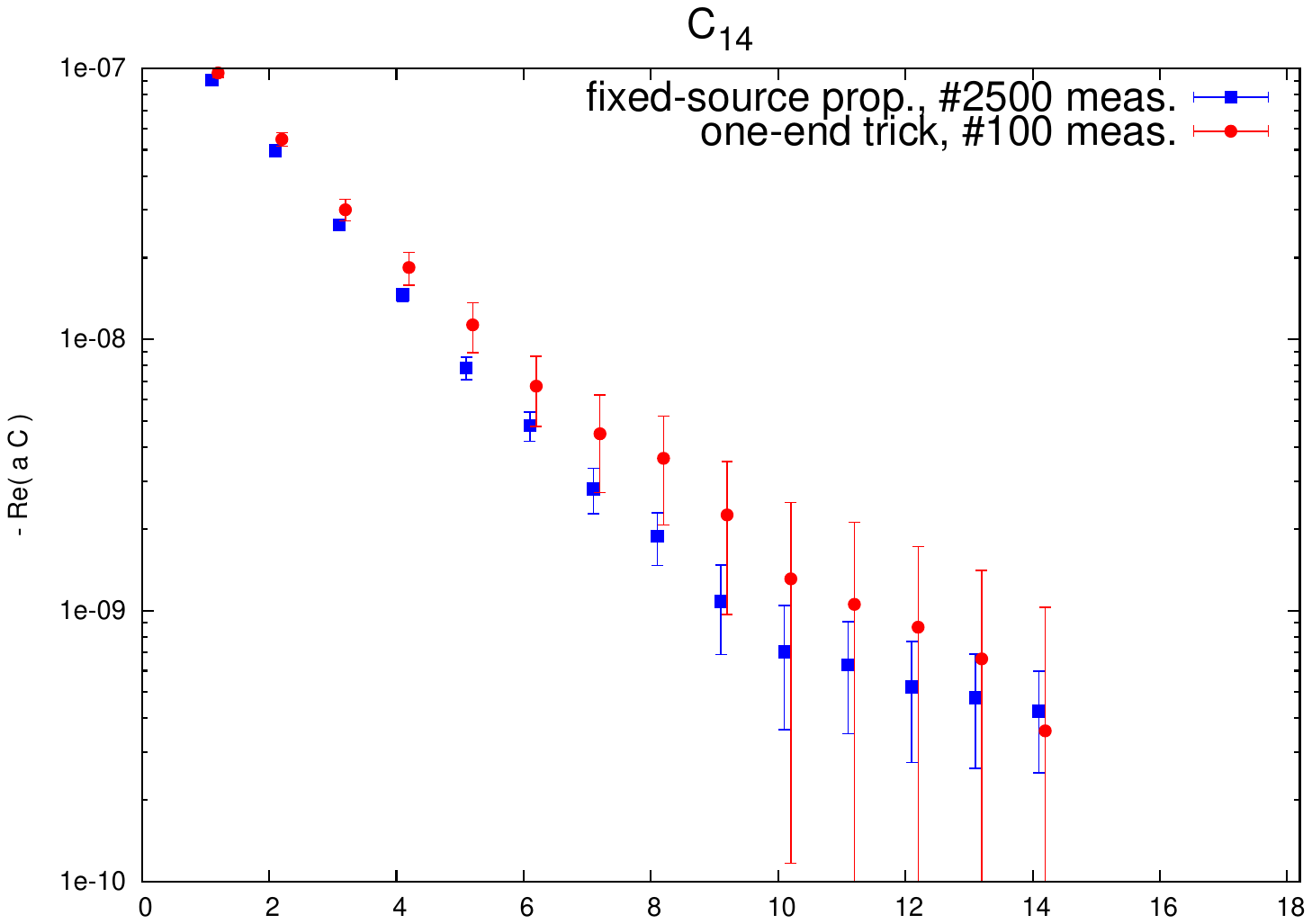}
\end{center}
\caption{\label{fig:psvsoe}Numerical comparison of methods for $C_{j 1}$, $j=1,2,3,4$.}
\end{figure}

In the following we demonstrate that different methods might yield significantly different signal-to-noise ratios. To this end we compute the four correlation matrix elements $C_{j 1}$, $j=1,2,3,4$. For $C_{1 1}$ we use
\begin{itemize}
\item[(1)] two fixed-source propagators (blue points in Fig.~\ref{fig:psvsoe}a),

\item[(2)] the one-end trick (red points in Fig.~\ref{fig:psvsoe}a).
\end{itemize}
For $C_{j 1}$, $j=2,3,4$ we use
\begin{itemize}
\item[(1)] two fixed-source propagators (fixed source, where only two propagators join) and a stochastic timeslice-to-all propagator for the closed fermion loop (blue points in Fig.~\ref{fig:psvsoe}b to Fig.~\ref{fig:psvsoe}d),

\item[(2)] the one-end trick and a stochastic timeslice-to-all propagator for the closed fermion loop (red points in Fig.~\ref{fig:psvsoe}b to Fig.~\ref{fig:psvsoe}d).
\end{itemize}

For all four diagrams the statistical errors obtained with methods (2) are roughly twice as large as those obtained with methods (1). Note, however, that for the latter the number of samples is larger by the factor $25$ (2,500 samples compared to 100 samples). Moreover, for each sample in (1) $12$ sources for light $u/d$ propagators had to be inverted, while for (2) $2$ inversion were sufficient (inversions for $s$ propagators are comparably cheap and, therefore, not taken into account). Consequently, at roughly fixed computational costs (2) yields a signal-to-noise ratio, which is larger by a factor $\approx \sqrt{25 \times 6} / 2 \approx 6$ compared to (1).

\section{\label{SEC002}Numerical results}


\subsection{\label{SEC003}Lattice setup}

\phantom{\speaker{Abdou Abdel-Rehim}}Using the methods described in the previous sections we have analyzed \cite{Edwards:2004sx} $N_f=2+1$ Wilson clover gauge link configurations generated by the PACS-CS collaboration \cite{Aoki:2008sm}. The lattice size is $32^3 \times 64$, the lattice spacing $a \approx 0.09 \, \textrm{fm}$. We have considered two ensembles corresponding to $m_\pi \approx 300 \, \textrm{MeV}$ (500 gauge link configurations; referred to as ``ensemble-A'') and $m_\pi \approx 150 \ \textrm{MeV}$ (198 gauge link configurations; referred to as ``ensemble-B'').

To improve the overlap with low lying states generated by our interpolating operators (\ref{eq:operatorone}) to (\ref{eq:operatorsix}), Gaussian smeared quark fields with APE smeared links are used. For ensemble-A we average each diagram $C_{j k}$, $j,k=1,2,3,4$ over 5 different source locations for each gauge link configuration, while for ensemble-B such an averaging has not been done yet. When there is a single closed fermion loop in a diagram, it is estimated with a stochastic timeslice-to-all propagator. When there are two closed fermion loops, we use a stochastic timeslice-to-all propagator for one of them and a fixed-source propagator for the other (cf.\ \cite{Wagner:2012ay} for a detailed discussion). Propagators connecting the two timeslices are also fixed-source propagators (i.e.\ we use method (1) from subsection~\ref{SEC010}).

We have extracted effective masses and energy levels from the $4 \times 4$ submatrix corresponding to interpolating operators (\ref{eq:operatorone}) to (\ref{eq:operatorfour}) by solving a standard generalized eigenvalue problem, with reference time $t_r = 1$,
\begin{eqnarray}
C(t) v_n(t,t_r) = \lambda_n(t,t_r) C(t_r) v_n(t,t_r) \quad , \quad E_n \overset{t \textrm{ large}}{=} E_n^\textrm{eff}(t,t_r) = \frac{1}{a} \ln\bigg(\frac{\lambda_n(t,t_r)}{\lambda_n(t+a, t_r)}\bigg) .
\end{eqnarray}


\subsection{Results for ensemble-A}

For ensemble-A our current results have smaller statistical errors than for ensemble-B, because of a larger number of samples (2,500 compared to 198) and the heavier pion ($m_\pi \approx 300 \, \textrm{MeV}$).


\subsubsection{Ignoring diagrams with closed fermion loops}

\begin{figure}[htb]
  \begin{center}
    \includegraphics[height=0.3\textwidth]{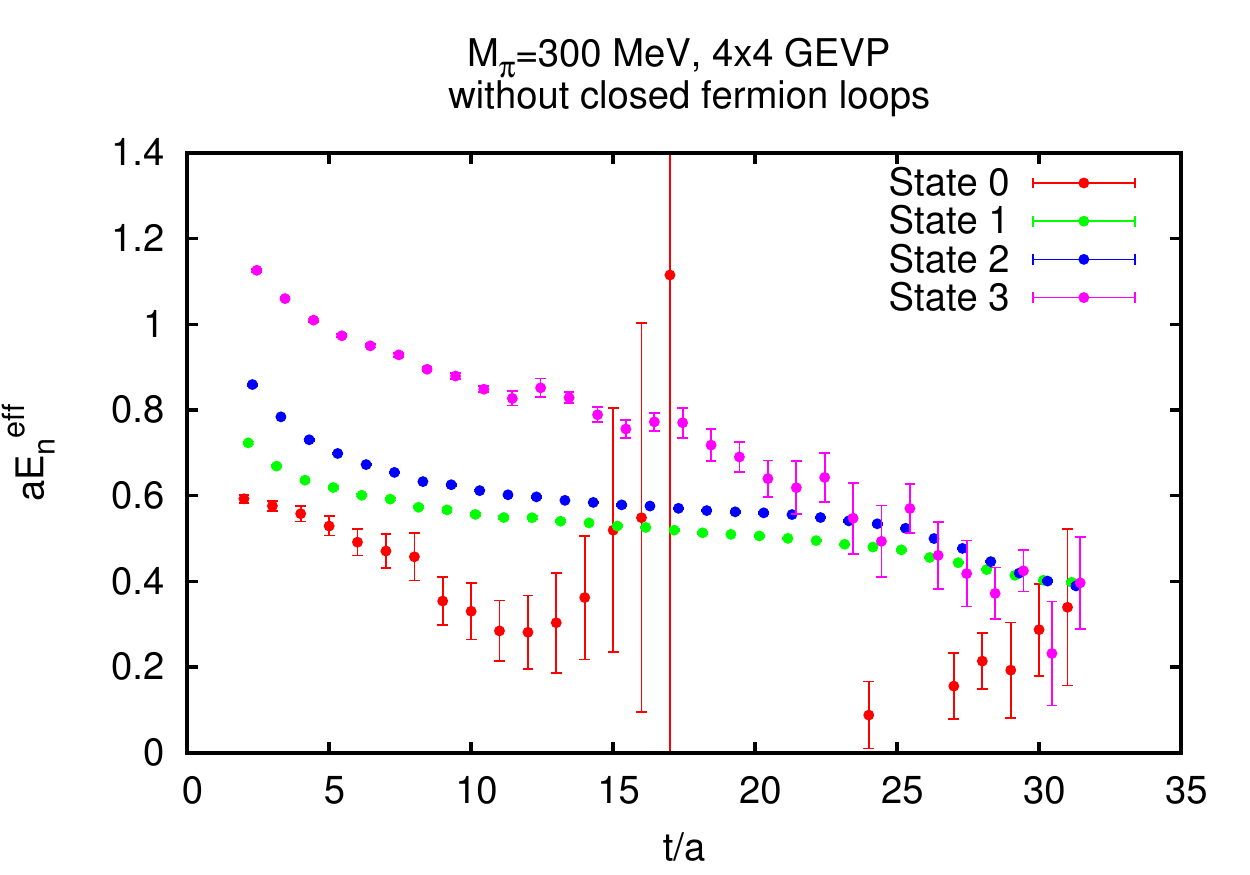}
    \includegraphics[height=0.3\textwidth]{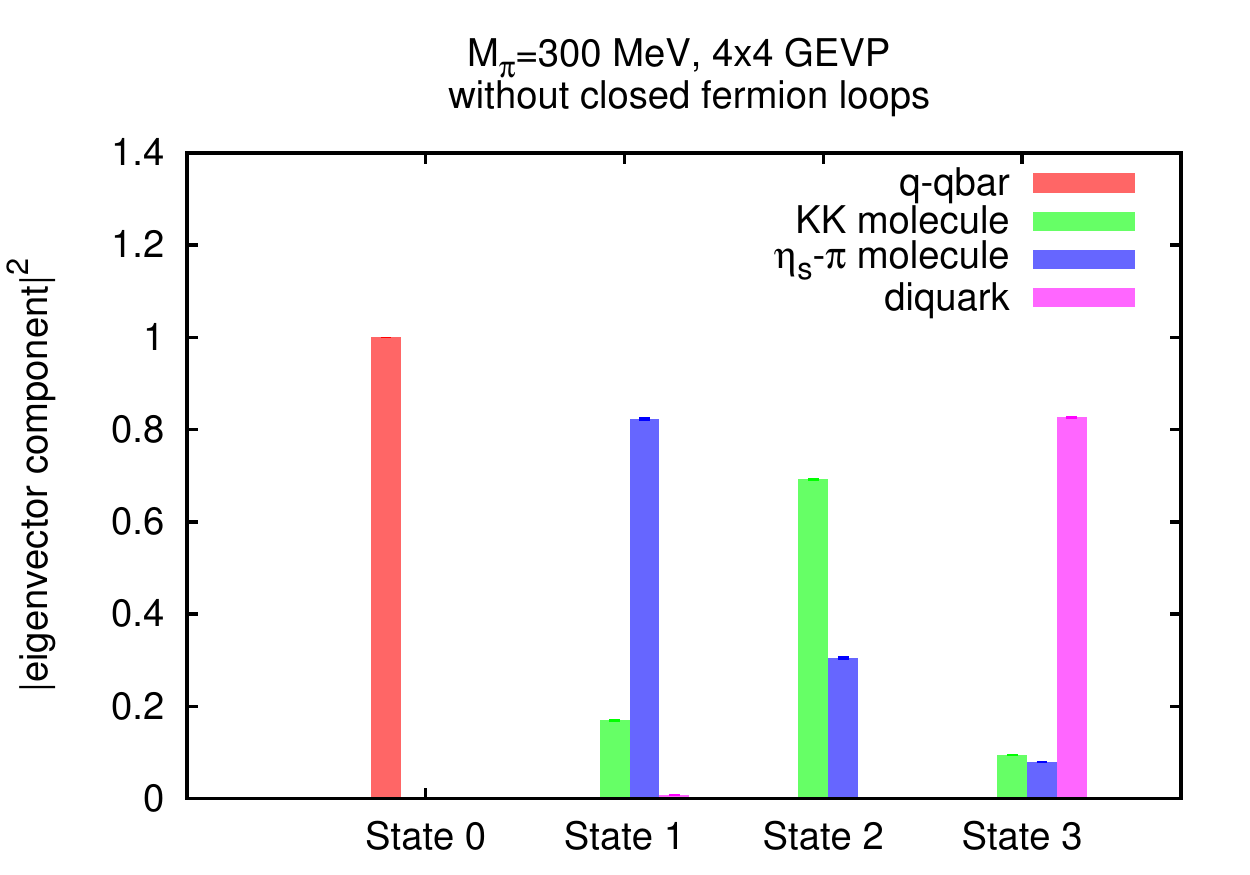}
   \end{center}
\caption{\label{300MeV_conn}Effective masses (left) and eigenvector components (right) for ensemble-A ignoring diagrams with closed fermion loops.}
\end{figure}

In Fig.~\ref{300MeV_conn} we show results ignoring diagrams with closed fermion loops (statistical errors are then significantly smaller). The $4 \times 4$ correlation matrix is then equivalent to a $1 \times 1$ matrix corresponding to the quark antiquark interpolator \eqref{eq:operatorone} and an independent $3 \times 3$ matrix corresponding to the four-quark bound state interpolators \eqref{eq:operatortwo} to \eqref{eq:operatorfour}.

The effective mass corresponding to the $1 \times 1$ matrix indicates an energy level in the region of the $a_0(980)$ ($980 \, \textrm{MeV} \times a \approx 0.45$). The statistical errors, however, are quite large, which might be an indication that this state is not predominantly of quark-antiquark type.

The lowest state extracted from the $3 \times 3$ matrix is dominated by a $\eta_{s} \pi$ interpolator and the second lowest state by the $K \bar{K}$ interpolator (cf.\ the right plot in Fig.~\ref{300MeV_conn} showing the eigenvector components $|v_n|^2$). This is in agreement with our recent study of the $a_0(980)$ using Wilson twisted mass quarks \cite{Daldrop:2012sr,Alexandrou:2012rm,Wagner:2012ay,Wagner:2013nta,Wagner:2013jda,Wagner:2013vaa}.

\subsubsection{Including diagrams with closed fermion loops}

\begin{figure}[htb]
  \begin{center}
    \includegraphics[height=0.3\textwidth]{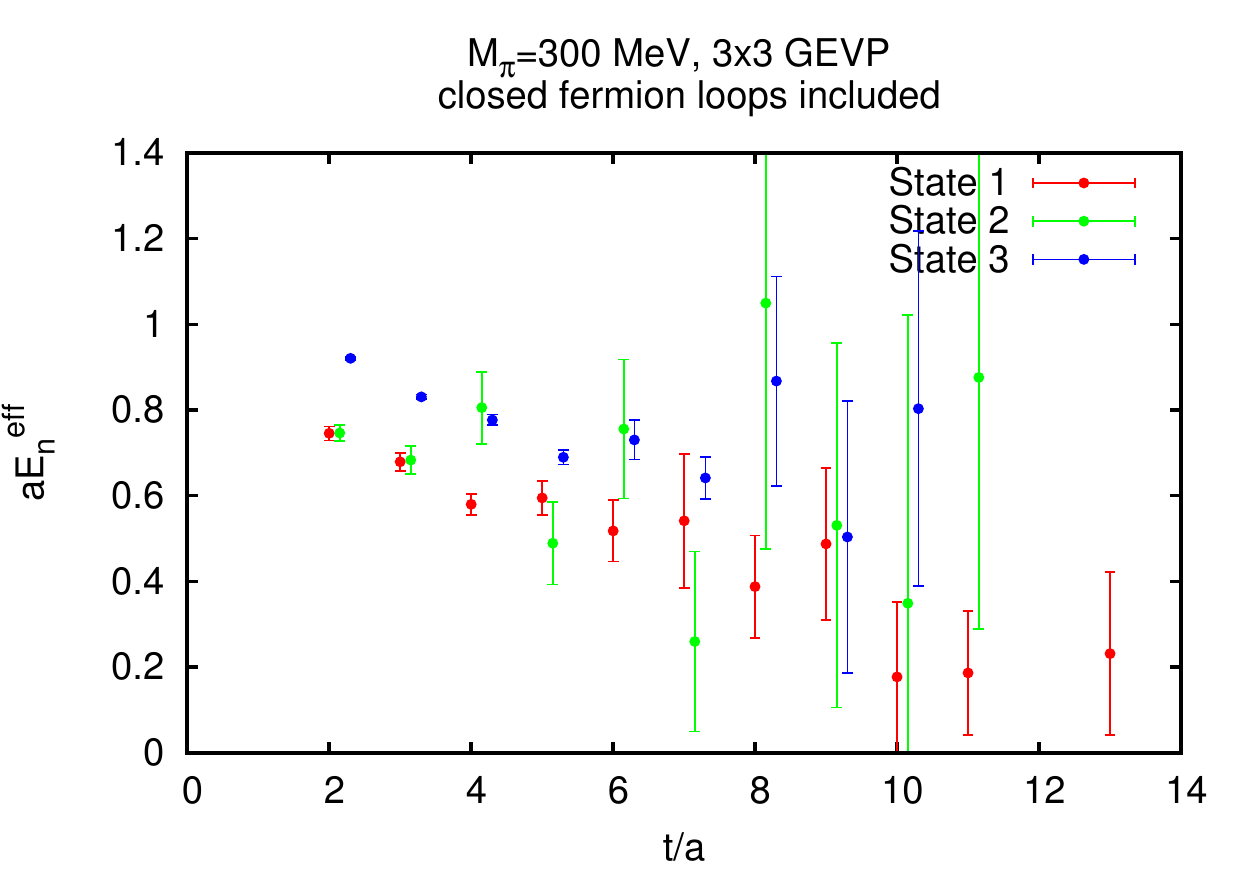}
    \includegraphics[height=0.3\textwidth]{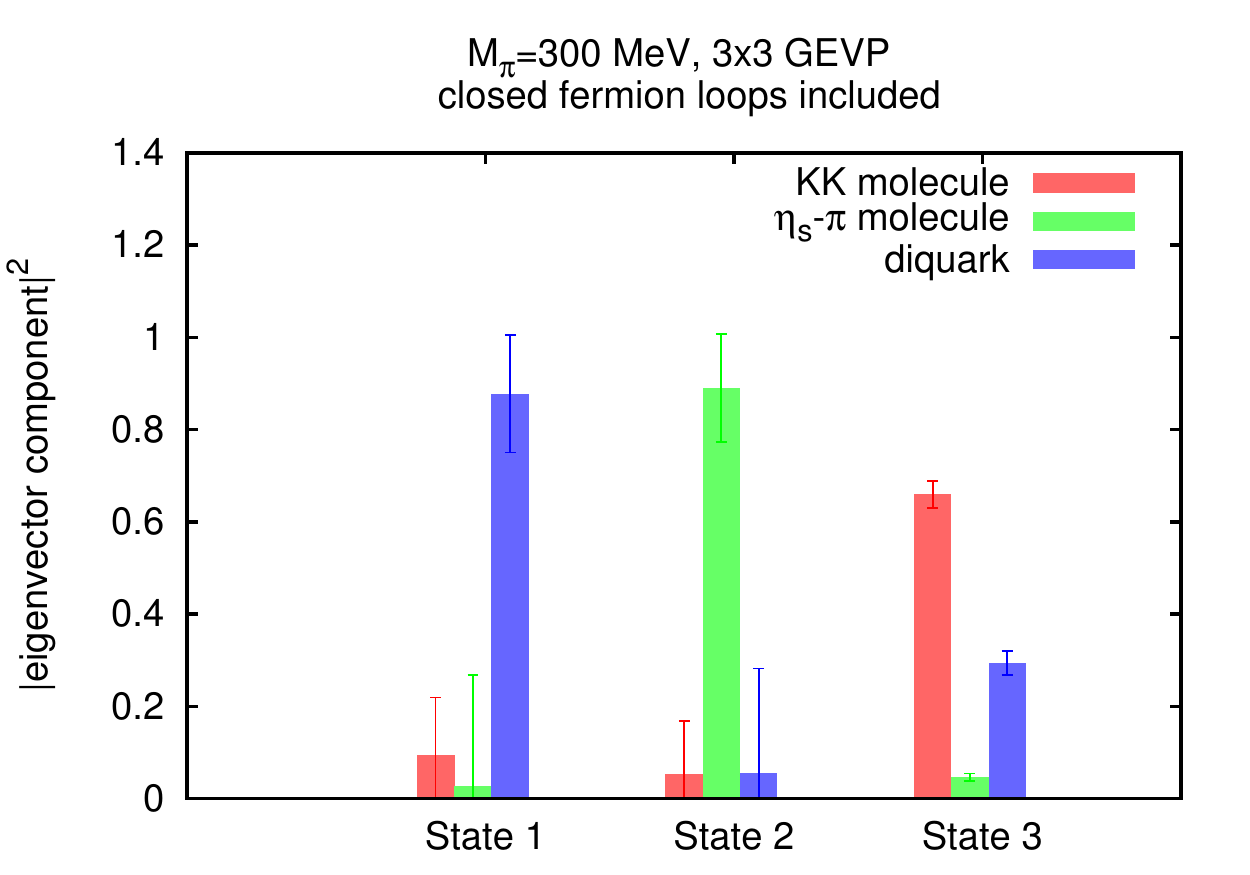}
    \includegraphics[height=0.3\textwidth]{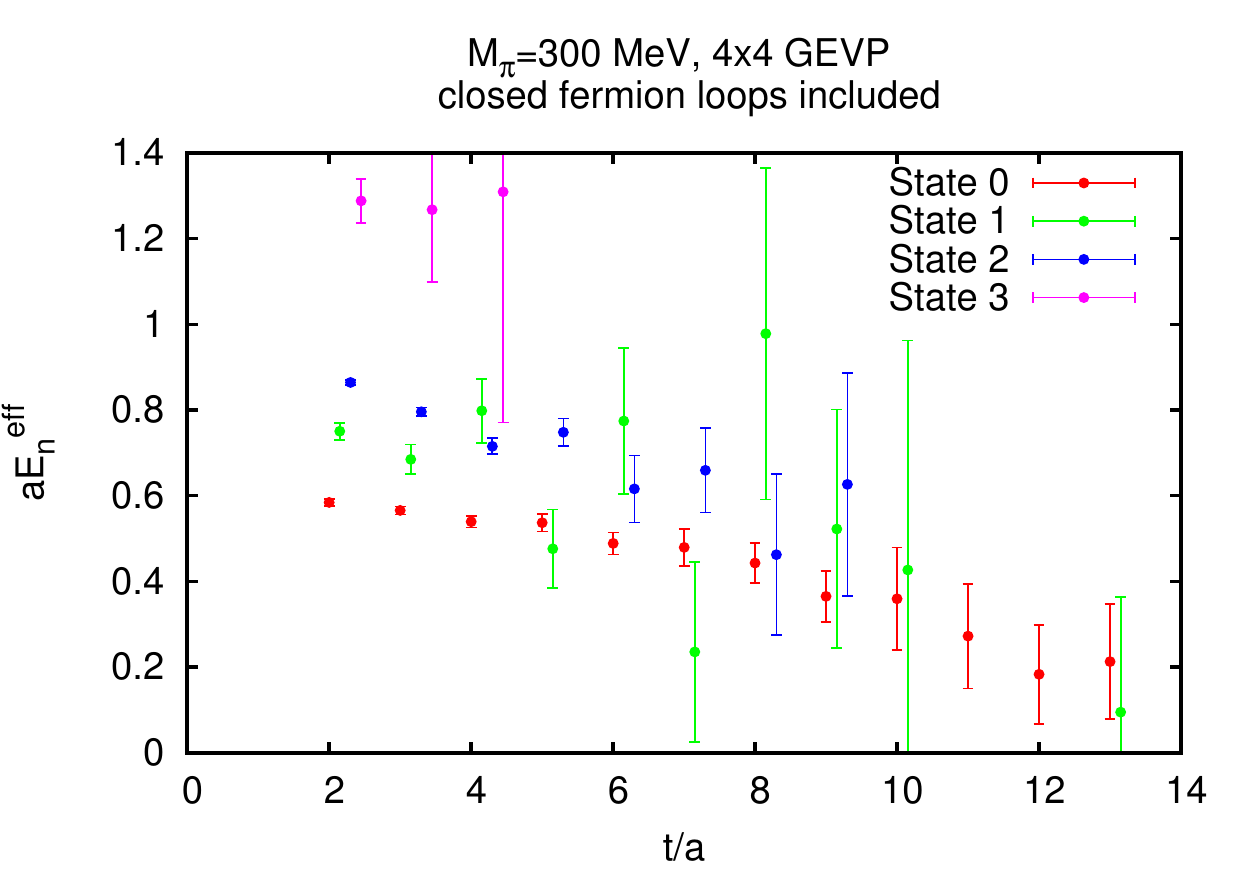}
    \includegraphics[height=0.3\textwidth]{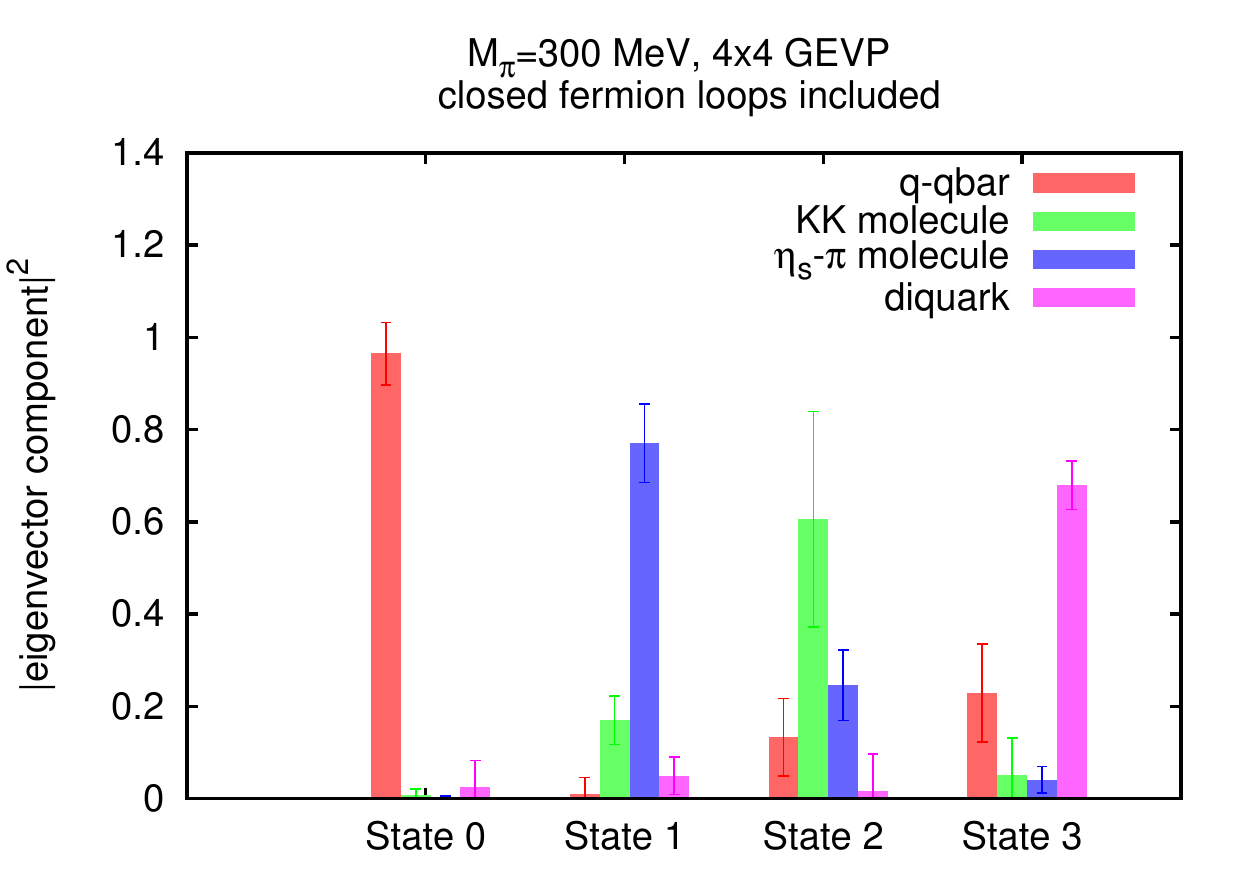}
   \end{center}
\caption{\label{300MeV_total}Effective masses (left) and eigenvector components (right) for ensemble-A including diagrams with closed fermion loops.}
\end{figure}

In Fig.~\ref{300MeV_total} we show results, where also diagrams with closed fermion loops are included. It is obvious that these diagrams significantly increase statistical errors. The plots in the first row correspond to a $3 \times 3$ matrix, where only the four-quark bound state interpolators \eqref{eq:operatortwo} to \eqref{eq:operatorfour} are included. The results shown in the second row were obtained from a $4 \times 4$ matrix, where also the quark-antiquark interpolator \eqref{eq:operatorone} has been considered. Note that, when taking closed fermion loops into account, the $4 \times 4$ matrix cannot be decomposed into independent $1 \times 1$ and $3 \times 3$ matrices.

The lowest state extracted from the $3 \times 3$ matrix is dominated by the diquark-antidiquark interpolator, which is in strong qualitative discrepancy to the $3 \times 3$ result obtained without closed fermion loops. The first and second excitation are of $\eta_{s} \pi$ and of $K \bar{K}$ type and seem to correspond to the lowest two states, when closed fermion loops are ignored. When advancing to the $4 \times 4$ matrix, the nature of the ground state changes from diquark-antidiquark to quark-antiquark type. The first and second excitation remain similar. These observations might be an indication that there is an additional state (besides $\eta_{s} + \pi$ and $K + \bar{K}$ two-meson states) in the mass region of the $a_0(980)$. This state seems to be more quark-antiquark-like than diquark-antidiquark-like.

Since at the moment the statistical errors are quite large, this interpretation is ambiguous. To clarify the situation and to possibly resolve a bound $a_0(980)$ state, we are currently extending the correlation matrix to include the two-meson interpolators \eqref{eq:operatorfive} to \eqref{eq:operatorsix}, which are known to result in less noisy correlators \cite{Alexandrou:2012rm}.


\subsection{Results for ensemble-B}

\begin{figure}[htb]
  \begin{center}
    \includegraphics[height=0.3\textwidth]{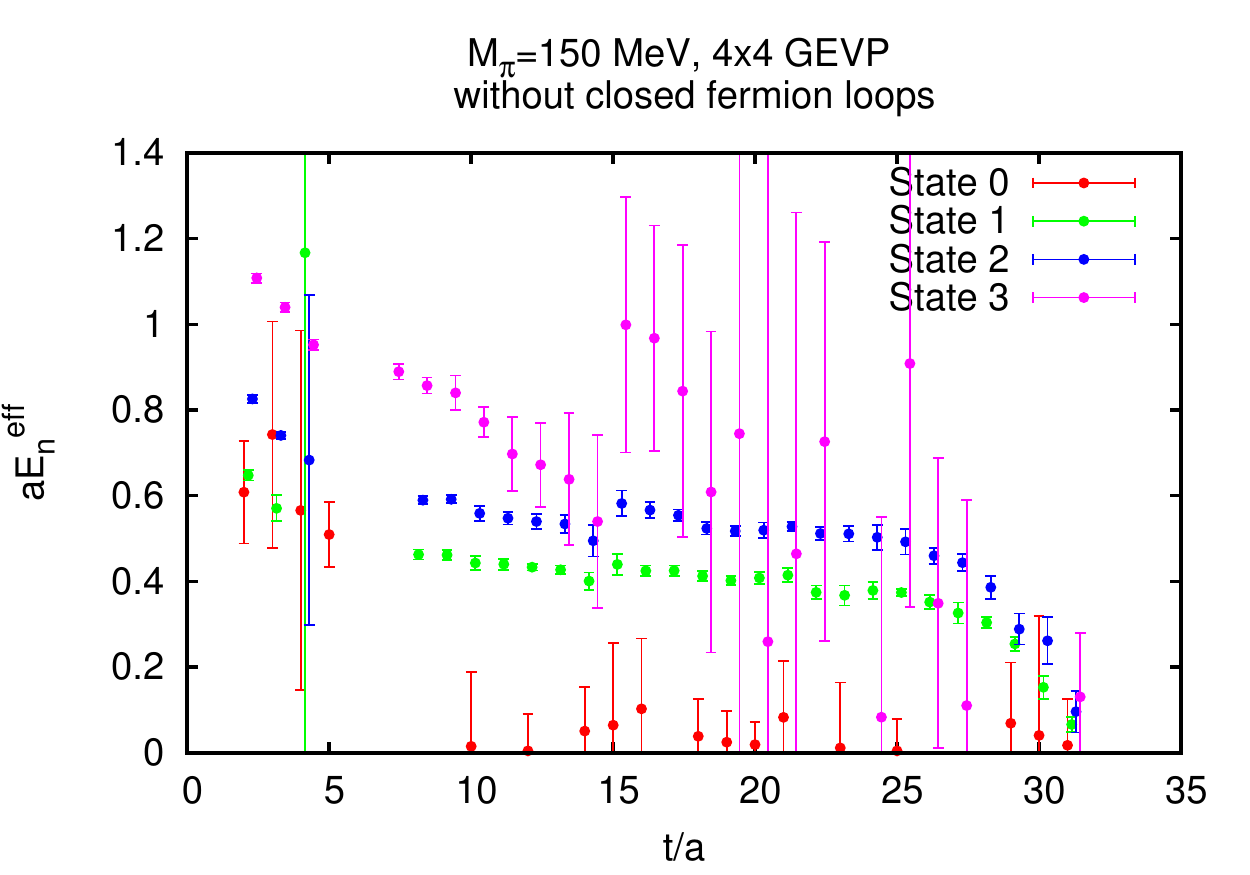}
    \includegraphics[height=0.3\textwidth]{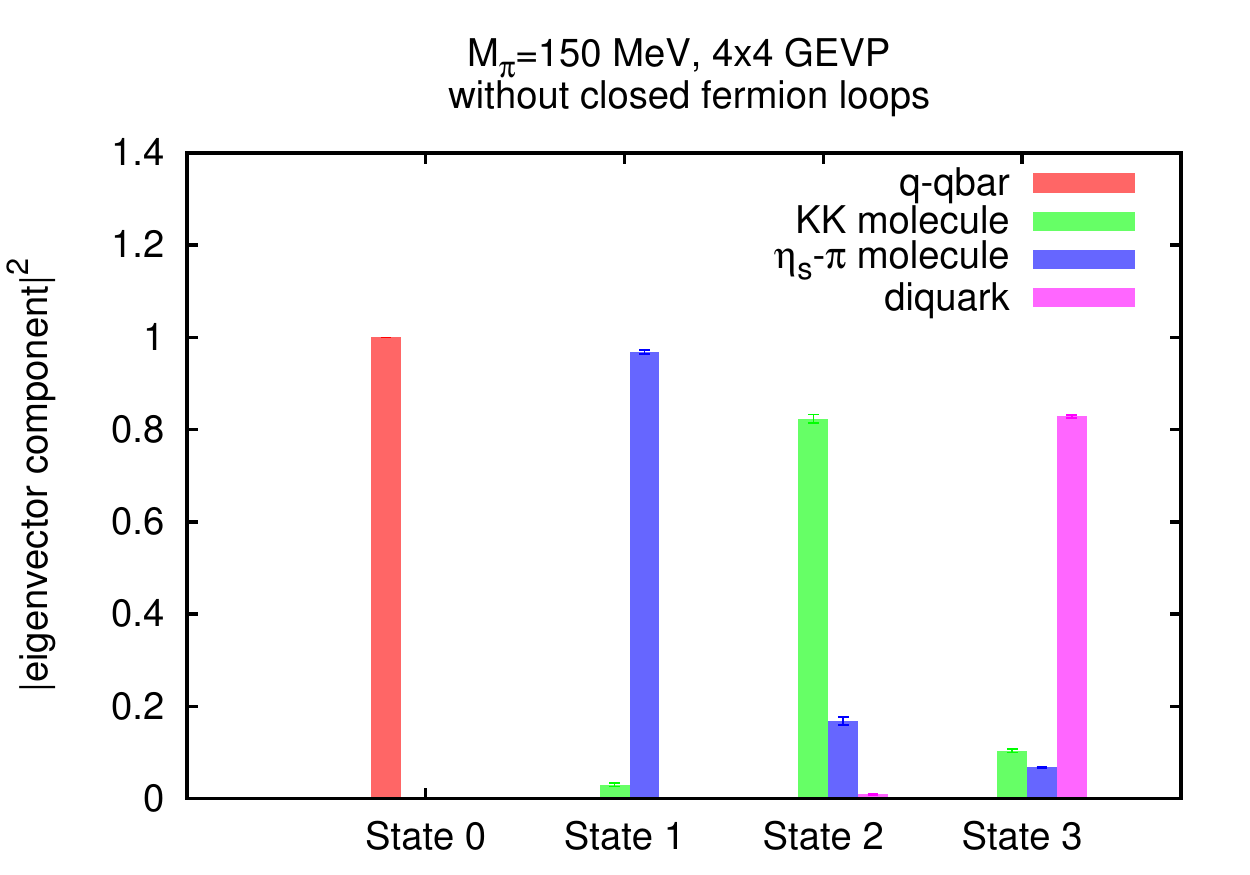}
   \end{center}
\caption{\label{150MeV_conn}Effective masses (left) and eigenvector components (right) for ensemble-A ignoring diagrams with closed fermion loops.}
\end{figure}

\begin{figure}[htb]
  \begin{center}
    \includegraphics[height=0.3\textwidth]{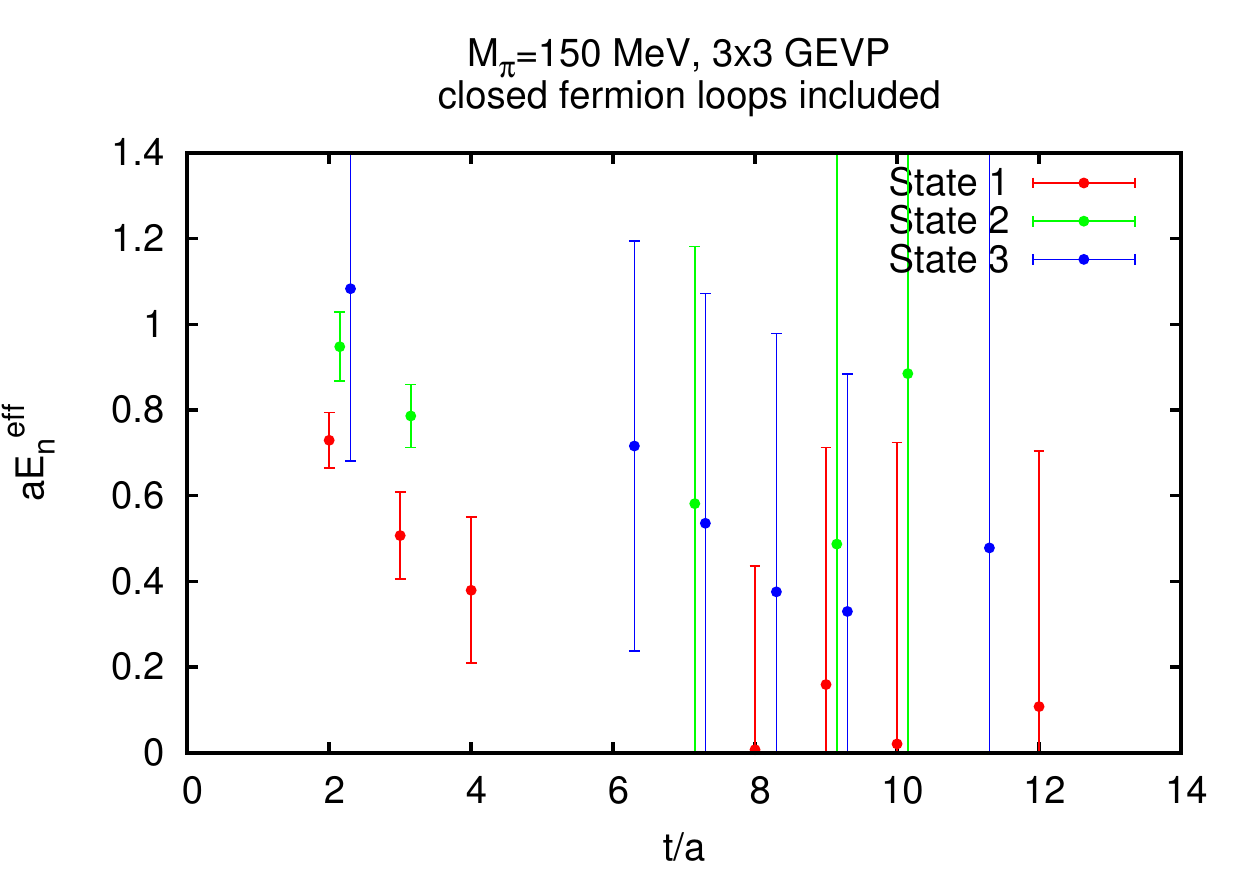}
    \includegraphics[height=0.3\textwidth]{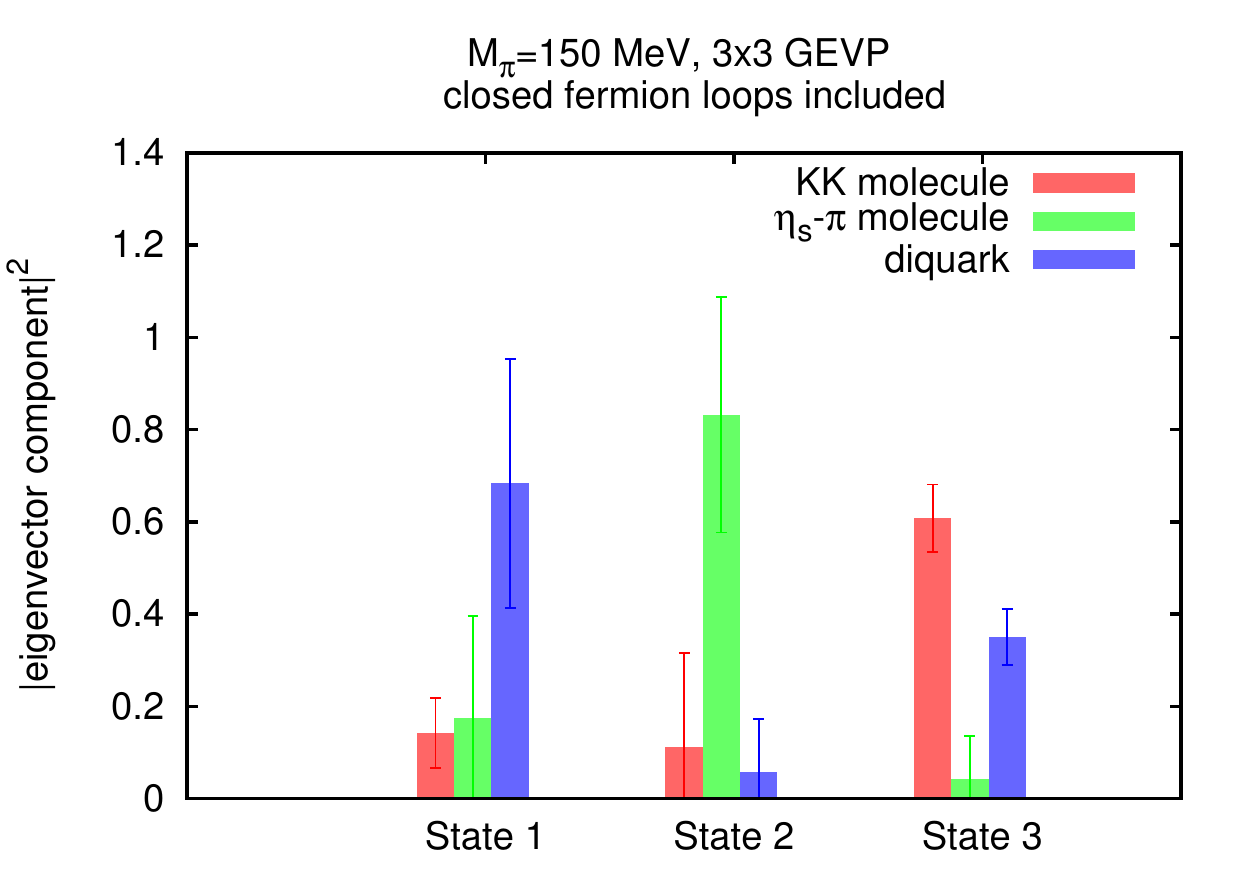}
    \includegraphics[height=0.3\textwidth]{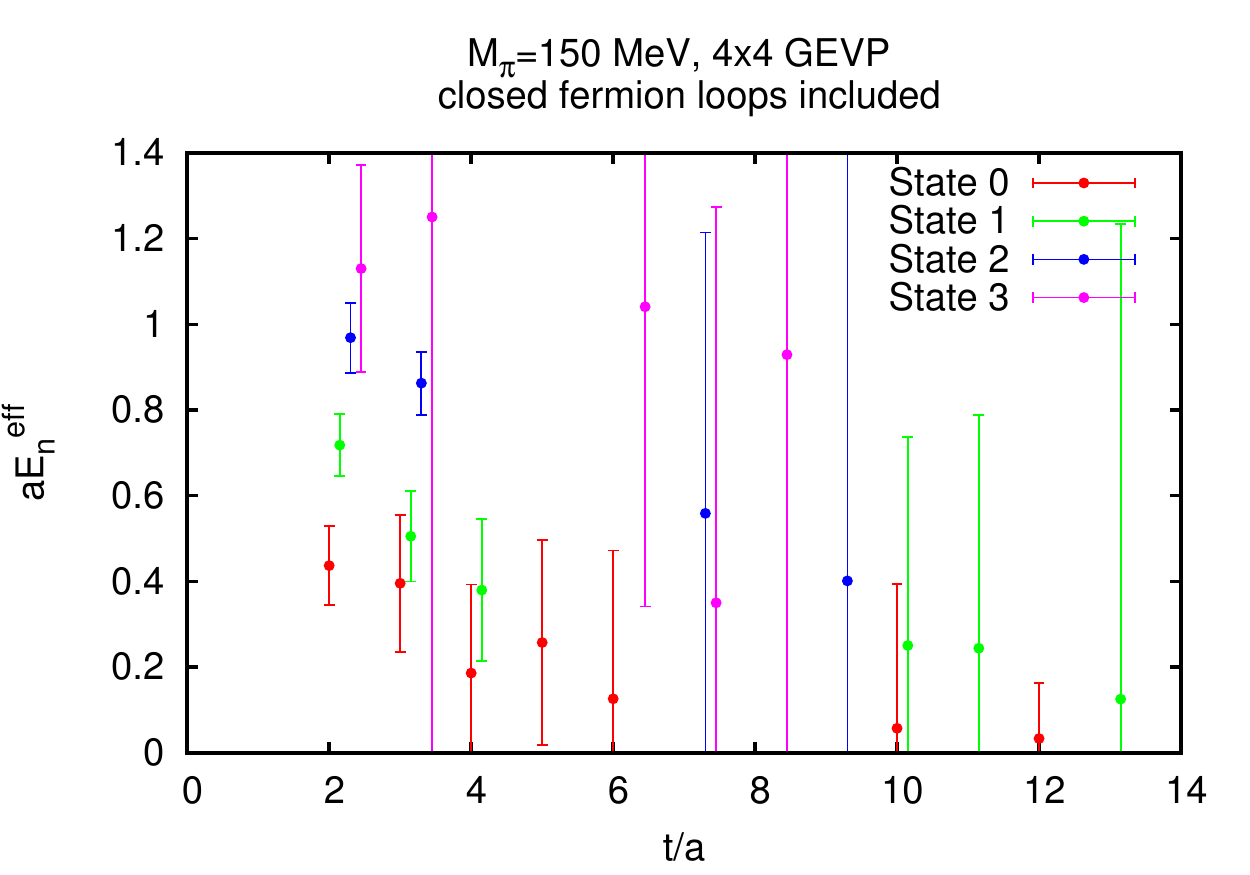}
    \includegraphics[height=0.3\textwidth]{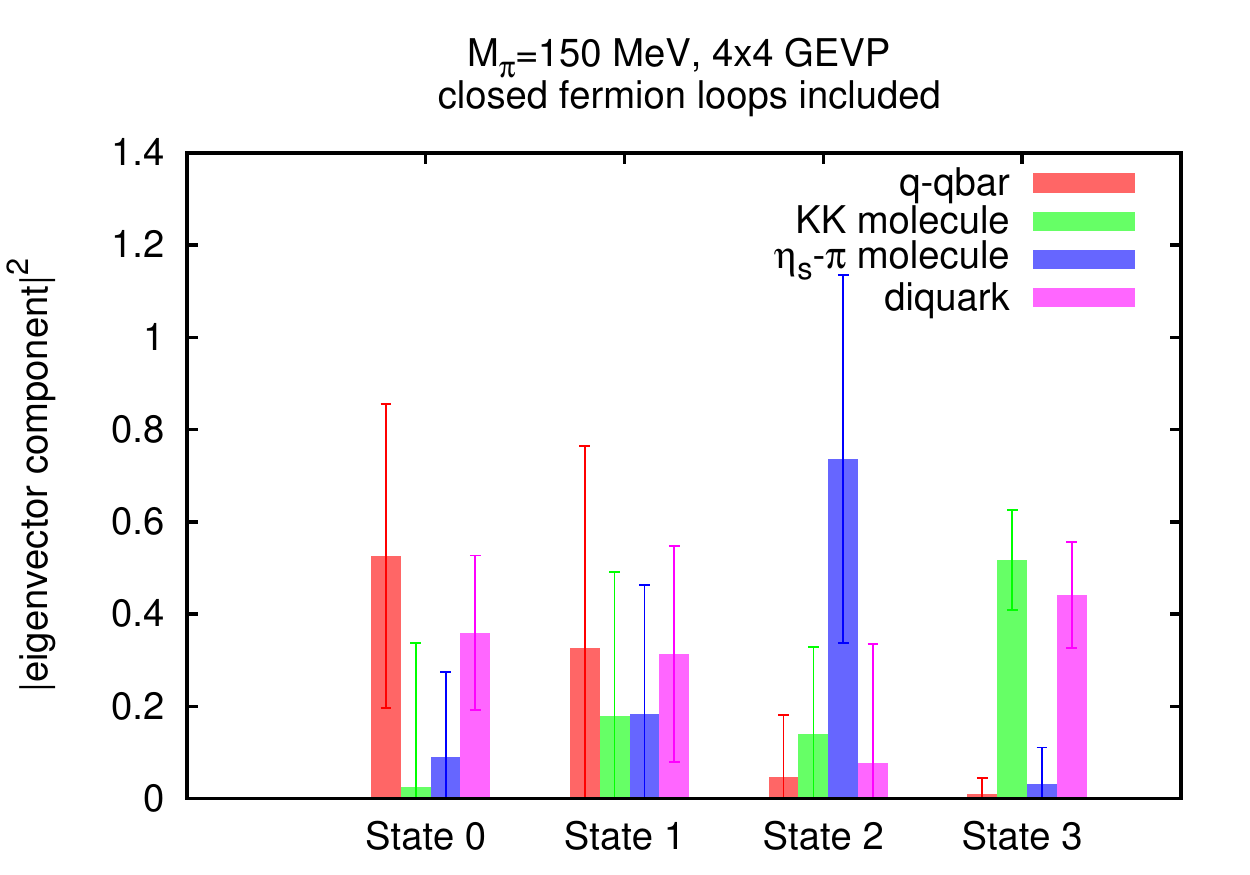}
   \end{center}
\caption{\label{150MeV_total}Effective masses (left) and eigenvector components (right) for ensemble-A including diagrams with closed fermion loops.}
\end{figure}

The results obtained for ensemble-B are shown in Fig.~\ref{150MeV_conn} and Fig.~\ref{150MeV_total}. There are only 198 samples at the moment and, hence, the results are more noisy, in particular when including closed fermion loops. The overall picture, however, is consistent with the one obtained for ensemble-A with respect to the operator content of the extracted states.

\section{Conclusions and outlook}

We presented technical aspects and preliminary results of our lattice QCD study of the scalar meson $a_0(980)$ using a variety of interpolating operators, e.g.\ of quark-antiquark, of mesonic molecule and of diquark-antidiquark type. Two $N_f=2+1$ ensembles of Wilson clover gauge link configurations have been analyzed, one at near physical pion mass $m_\pi \approx 150 \ \textrm{MeV}$. Contributions from closed fermion loops, which are technically challenging to compute, have been taken into account.

Our main goal is to identify and study the $a_0(980)$ state and understand its quark sub-structure. Our current results indicate that including quark-antiquark interpolators as well as diagrams with closed fermion loops have an important effect on the extracted spectrum. Currently we are working on reducing the statistical errors, which we plan to achieve in two ways: first, we will include explicitly scattering states of $\eta_s + \pi$ and $K + \bar{K}$ type; second we implement more advanced combinations of techniques, in particular making extensive use of the one-end trick.


\begin{acknowledgments}

M.D.B.\ is funded by the Irish Research Council and is grateful for the hospitality at the University of Cyprus, the Cyprus Institute, and DESY Zeuthen, where part of this work was carried out. 
The work of M.G.\ was supported by the European Commission, European Social Fund and Calabria Region, that disclaim any liability for the use that can be done of the information provided in this paper.
M.W.\ acknowledges support by the Emmy Noether Programme of the DFG (German Research Foundation), grant WA 3000/1-1.
This work was supported in part by the Helmholtz International Center for FAIR within the framework of the LOEWE program launched by the State of Hesse.

\end{acknowledgments}



\end{document}